\documentclass[aps,prl,reprint,superscriptaddress]{revtex4-2}
\usepackage{amsmath,amssymb,graphicx}
\usepackage{xcolor}

\begin{document}

\title{Nuclear Heterodyne Interferometry for Gravitational Spectroscopy}

\author{Ralf Röhlsberger}
\email{r.roehlsberger@hi-jena.gsi.de}
\affiliation{Helmholtz Institut Jena, Fraunhoferstr. 8, 07743 Jena, Germany}
\affiliation{GSI Helmholtz Institut f\"ur Schwerionenforschung, Planckstr. 1, 64291 Darmstadt, Germany}
\affiliation{Institut f\"ur Optik und Quantenelektronik, Friedrich-Schiller-Universit\"at Jena, Max-Wien Platz 1, 07743 Jena, Germany}
\affiliation{Deutsches Elektronen-Synchrotron DESY, Notkestr. 85, 22607 Hamburg, Germany}
\date{\today}

\begin{abstract} 
Gravitational spectroscopy tests the coupling of gravity to matter by measuring gravitationally induced frequency shifts of quantum transitions. While modern optical clocks probe the gravitational response of electronic transitions with extraordinary precision, tests in the nuclear sector have not progressed since the Mössbauer measurements of the gravitational redshift by Pound and Rebka. Here we introduce a new approach to nuclear gravitational spectroscopy based on phase-sensitive heterodyne interferometry of time-resolved nuclear resonant scattering of synchrotron radiation. In this scheme the gravitational redshift appears as a slowly accumulating phase drift of a delayed heterodyne beat signal, converting nuclear gravitational spectroscopy from energy-domain detection to time-domain interferometry. A Fisher-information analysis supported by Monte Carlo simulations and experimentally confirmed photon count rates shows that the nuclear gravitational redshift of $^{57}$Fe can be detected within hours on a few-meter-scale vertical baseline, with percent-level precision on deviations from general relativity becoming accessible on day-scale timescales. The method thus establishes an experimentally realistic and scalable platform for precision tests of gravitational coupling to nuclear structure.
\end{abstract}

\maketitle

\textit{Introduction} -- 
Tests of the gravitational redshift (GRS) provide one of the most direct
experimental probes of general relativity and the Einstein
equivalence principle, spanning laboratory and spaceborne clock
experiments, modern optical frequency metrology, astrophysical
observations, and global navigation systems
\cite{Vessot1980b,Chou2010,McGrew2018,Herrmann2018,Greenstein1971,Will2014}.
Nuclear gamma-ray spectroscopy enabled the first laboratory tests of the gravitational redshift through the $^{57}$Fe Mössbauer
experiments of Pound and Rebka \cite{PoundRebka1960,PoundRebka1960PR}. Later refinements by
Pound and Snider \cite{PoundSnider1964,PoundSnider1965}, and the measurements by Potzel et al. using
$^{67}$Zn \cite{Potzel1994,Potzel1995,Schiessl1996} confirmed the relativistic prediction at the percent level.
Modern optical clocks now probe gravitational frequency shifts with
extraordinary precision and have established gravitational spectroscopy as a powerful technique for testing gravity \cite{Chou2010,Nicholson2015,McGrew2018,Bothwell2022,Lisdat2016,Derevianko2014,Takamoto2020}.
Despite these advances, experimental tests in the nuclear sector have not progressed since the early Mössbauer measurements. Nuclear transitions probe a regime complementary to optical clocks because their transition energies originate predominantly from nuclear binding energies governed by the strong interaction rather than electronic structure, potentially providing sensitivity to composition-dependent deviations from general relativity \cite{Nordtvedt1968,Flambaum2007,Flambaum2006,Safronova2018,Arvanitaki2015}.

Over the past three decades nuclear resonant scattering with
synchrotron radiation has become a powerful method for
time-resolved Mössbauer spectroscopy.
Synchrotron radiation provides high spectral brilliance
and a pulsed time structure that allows direct observation
of the delayed nuclear response \cite{Gerdau1985,Hastings1991,Alp1993,Sturhahn1998,GerdauDeWaard1999,Shvydko1999,Roehlsberger2004,Shenoy2008,Roehlsberger2010,Roehlsberger2012}
and enables phase-sensitive detection techniques \cite{Shvydko1994,Callens2002,Heeg2015,Heeg2021Coherent,Bocklage2021Coherent,Velten2024,Yuan2025,Negi2025Energy}
not accessible in conventional
energy-domain Mössbauer spectroscopy.

\begin{figure*}[t]
\centering
\includegraphics[width=\linewidth]{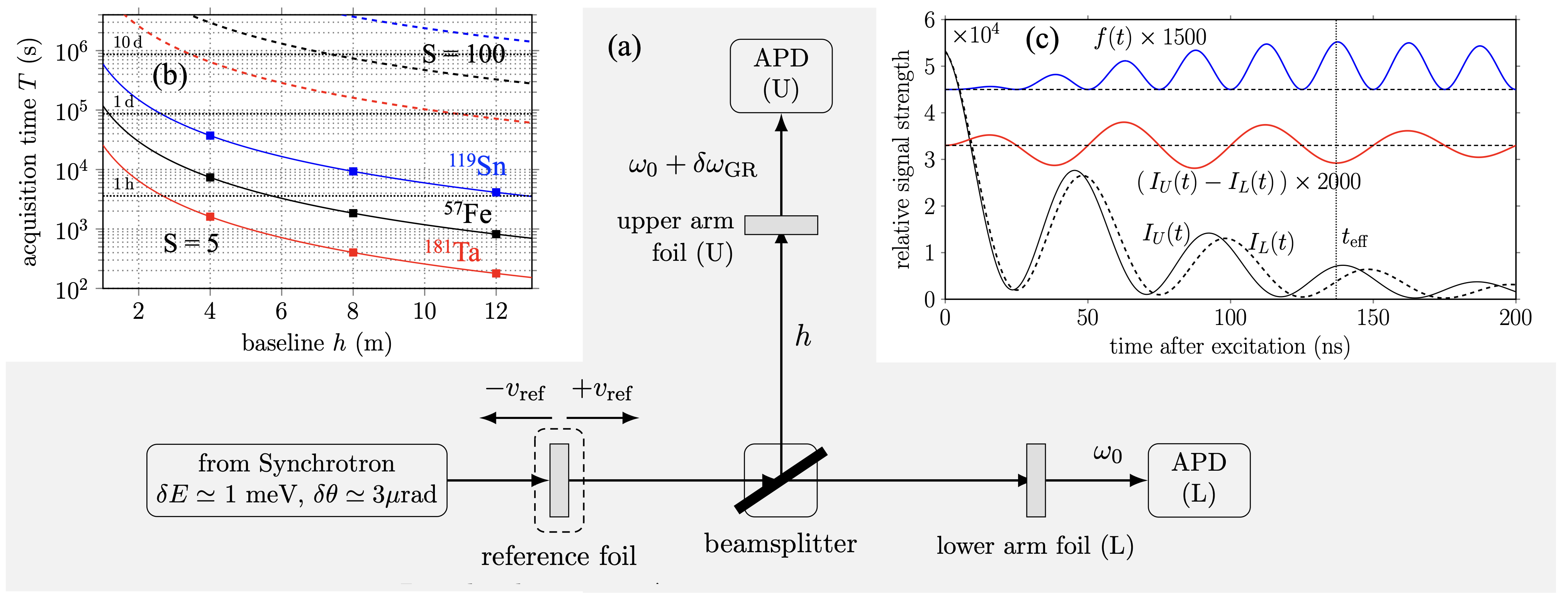}
\caption{Concept of nuclear heterodyne interferometry for gravitational spectroscopy. 
(a)  A Doppler-driven single-line reference absorber introduces a small heterodyne detuning to an incident monochromatic X-ray pulse. After a crystal beamsplitter the radiation interacts with two identical single-line nuclear absorbers separated by a vertical baseline $h$. The delayed nuclear responses form heterodyne beat signals whose relative phase evolves due to the gravitational redshift, see main text for a detailed description. 
(b) Acquisition times for $S$ = 5 (solid lines) and $S = 100$ (dashed lines) as a function of vertical baseline~$h$, computed from Eq.\,(\ref{Fisher02}) for all three benchmark isotopes. Filled squares mark the Monte Carlo survey values from \cite{Supplemental}, Table~V. Parameters: $C = 0.9$.
(c) Delayed heterodyne signals $I_U(t)$ and $I_L(t)$ for the upper and lower interferometer arms for the $^{57}$Fe example discussed in the text after Eq.\,(\ref{Fisher02}). The heterodyne phase drift induced by the GRS in the upper arm is strongly magnified here for visualization. The differential signal is shown in red with a vertical offset and scale factor 2000; the scaled Fisher-information density $f(t)$ is plotted in blue, with the maximum of the envelope defining the effective phase accumulation time $t_{\rm eff}$}
\label{Fig1}
\end{figure*}

Here we propose a new approach to nuclear GRS measurements
based on phase-sensitive nuclear heterodyne interferometry
using time-resolved nuclear resonant scattering of
synchrotron radiation. In this scheme the gravitational
frequency shift $\delta\omega$ appears as a slowly accumulating phase
difference between two heterodyne signals recorded in
spatially separated interferometer arms. In this sense, nuclear heterodyne interferometry realizes the
nuclear analogue of phase-sensitive interferometric spectroscopy
known from Ramsey interrogation of atomic clocks \cite{Ramsey1950,RamseySilsbee1951,Ramsey1990, Ludlow2015}. Because the full
delayed photon waveform contributes to the statistical
inference of the frequency shift, our method provides
substantially higher sensitivity than conventional
energy-domain detection. 

Using experimentally demonstrated photon fluxes in synchrotron Mössbauer spectroscopy \cite{Velten2024}, we develop in the following sections a complete statistical framework for nuclear heterodyne interferometry — from the Fisher-information analysis of the time-resolved waveforms to a systematic Monte Carlo validation across three isotopes and four baselines. The analysis shows that the method operates in a regime where the nuclear gravitational redshift is detectable within hours, and deviations from general relativity can be constrained at the percent level within standard synchrotron beamtimes, establishing a concrete experimental path toward improved nuclear-sector tests of the equivalence principle; see \cite{Supplemental}, Sec.\,II.

\textit{Nuclear heterodyne interferometry} --
The central idea of nuclear heterodyne interferometry is to
convert the gravitational redshift between two single-line nuclear oscillators
located at different heights into a measurable phase drift
between them, observed in the time domain, as illustrated in Fig.\,\ref{Fig1}a.
A reference absorber placed in the incident beam is driven
with a controlled Doppler velocity, producing a frequency
detuning $\Delta\omega_{\rm het}$ relative to the resonant
transition. This detuning generates a delayed heterodyne
oscillation in the time-resolved nuclear scattering signal.
Because the heterodyne detuning is externally controlled, it
also provides a natural symmetry handle for separating the
gravitational signal from instrumental offsets. Reversing the
Doppler velocity of the reference absorber changes the sign of
the heterodyne detuning, $\Delta\omega_{\rm het}\to-\Delta\omega_{\rm het}$,
while leaving the gravitational shift $\delta\omega$ unchanged. Measurements
for opposite detunings therefore help separate instrumental
frequency offsets from the gravitational signal and can be
naturally interleaved during the forward and return motion of
the drive.
\paragraph{Heterodyne signals.}
After a 50:50 beamsplitter the radiation interacts with two
identical absorbers located at different gravitational
potentials. The transmitted photons in the upper and
lower arms are detected by fast avalanche photodiodes
that record the delayed nuclear response following each
synchrotron pulse. The delayed intensity signals recorded in the two interferometer
arms, in the following also referred to as waveforms, can be written as \cite{Supplemental}, Sec.\,III :
\begin{eqnarray}
\label{LowerArm}
I_L(t) & = & R\, e^{-\Gamma t} \left[1+C\cos(\Delta\omega_{\rm het} t)\right],\\[2mm]
\label{UpperArm}
I_U(t)& = & R\, e^{-\Gamma t} \left[1+C\cos\!\left((\Delta\omega_{\rm het}+\delta\omega)t\right)\right],
\end{eqnarray}
where \(R\) is the usable delayed count rate per arm, \( \Gamma=n\Gamma_0 \)
the effective linewidth (where $n$ accounts for the linewidth broadening of the effective single-line absorber, 
either due to superradiant effects or residual inhomogeneous broadening), and \(C\) the heterodyne contrast.
Eqs.\,(\ref{LowerArm}) and (\ref{UpperArm}) therefore describe two nuclear oscillators whose resonance
frequencies differ by the gravitational redshift $\delta\omega$.
Subtracting the two normalized and temporally aligned detector signals in Eqs.\,(\ref{LowerArm}) and (\ref{UpperArm}) yields
\begin{equation}
\label{ArmsDifference}
I_U(t)-I_L(t)
\simeq
- R\, e^{-\Gamma t} C\, \delta\omega\, t
\sin(\Delta\omega_{\rm het} t),
\end{equation}
showing that the gravitational redshift appears as a phase drift
$\delta\phi =\delta\omega\,t$ that increases linearly in time.
The differential signal is maximal near the steepest slopes
of the oscillation, providing the time-domain analogue of
the Pound--Rebka detection scheme.
Rather than sampling a narrow
region of an energy-domain resonance slope, the
gravitational signal is accumulated coherently across the
full delayed nuclear waveform.
The dual-arm geometry suppresses common-mode fluctuations
such as synchrotron timing jitter and global energy drifts,
which cancel in the differential signal (see \cite{Supplemental}, Sec.\,IV. A).

\paragraph{Extraction of the gravitational redshift.}
Efficient extraction of the gravitational phase drift
requires an appropriate heterodyne detuning
$\Delta\omega_{\rm het}$. The heterodyne detuning is chosen such that several oscillation
periods occur within the delayed observation window. Practically, this window is set by the useful 
delayed signal before either the count rate becomes negligible or the bunch-to-bunch distance
limits the usable time interval.

The waveforms $I_L(t)$ and $I_U(t)$ are normalized over an early-time reference window in which $\delta\omega\,t \ll 1$, yielding $\tilde{I}_{L,U}(t)$ with arm-dependent transmission and detection-efficiency factors removed. The gravitational frequency shift is then extracted from the differential waveform
\begin{equation}
\Delta I(t) = \tilde{I}_U(t) - \tilde{I}_L(t).
\end{equation}
Since \(\delta\omega \ll \Delta\omega_{\rm het}\) basically always holds,
the differential signal assumes the leading-order form \cite{Supplemental}, Sec.\,III\,:
\begin{equation}
\Delta I(t)=
\delta\omega\,K(t),
\label{eq:template_form}
\end{equation}
where the known temporal kernel is
\begin{equation}
K(t)=
- R\,C\,e^{-\Gamma t}\,
t\,\sin(\Delta\omega_{\rm het}t).
\label{eq:kernel}
\end{equation}
The gravitational frequency shift $\delta\omega$ therefore enters as the amplitude of a known waveform shape $K(t)$, so that, to leading order, the redshift can be obtained from a one-parameter fit to the measured differential signal $\Delta I(t)$. The waveform fit thereby yields directly the differential frequency $\delta\omega$ between the two interferometer arms.

Possible deviations from the relativistic prediction can
be parameterized by $\delta\omega = (1+\alpha)\,\delta\omega_{\rm GR}$,
where $\alpha=0$ corresponds to the prediction of general relativity.
The statistical uncertainty $\sigma_{\delta\omega}$ of the extracted parameter
$\delta\omega$ is governed by the Poisson counting statistics
of the delayed photon counts, which can be mapped to the corresponding uncertainty
$\sigma_\alpha$ of the deviation parameter $\alpha$, $\sigma_{\alpha} = \sigma_{\delta\omega} / \delta\omega_{\rm GR}$,
i.e., the precision on $\alpha$ corresponds to the fractional precision with which the gravitational frequency shift $\delta\omega$ is measured, as explained in the following paragraph by using the concept of Fisher information.  

\textit{Statistical sensitivity and Fisher information} --
For Poisson statistics the achievable precision for $\delta\omega$ is
naturally quantified by the Fisher information of the time-resolved signal \cite{Kay1993,VanTrees2001}.
\begin{equation}
\mathcal{F}(\delta\omega) = \int_0^T f(t)\,dt =
\int_0^T
\frac{1}{I_U(t)}
\left(
\frac{\partial I_U(t)}{\partial \delta\omega}
\right)^2 dt ,
\end{equation}
where $f(t)$ is the Fisher-information density.
For the heterodyne waveform $I_U(t)$, $\mathcal{F}(\delta\omega)$
yields the Cram\'er--Rao bound, i.e., the smallest statistical uncertainty achievable for  $\delta\omega$:
\begin{equation}
\label{sigmadelta}
\sigma_{\delta\omega} = \frac{1}{\sqrt{\mathcal{F}(\delta\omega)}}  \sim\frac{1}{C\,t_{\rm eff}\sqrt{R\,T}}.
\end{equation}
where $R$ is the usable delayed count rate, $C$ the heterodyne
contrast, and $T$ the data acquisition time. Because the phase drift grows linearly with time while the nuclear response decays as $e^{-\Gamma t}$, the Fisher-information density $f(t)$ scales approximately as $t^2 e^{-\Gamma t}$. Its maximum defines the characteristic phase-accumulation time $t_{\rm eff} = 2/\Gamma$ in the idealized limit; see Fig.\,1c and \cite{Supplemental}, Sec.\,V.

It is convenient to rewrite Eq.~(\ref{sigmadelta}) in terms of the significance parameter 
$S=\delta\omega / \sigma_{\delta\omega}$ that is, the signal size measured in units of 
its statistical uncertainty; in the Gaussian limit, this is equivalent to the usual signal-to-noise ratio.
Solving Eq.~(\ref{sigmadelta}) for the acquisition time $T$ to reach a target significance $S$ gives (see \cite{Supplemental}, Sec. VI)
\begin{equation}
T =
\frac{2\,S^2}{\beta R\,C^2\,t_{\mathrm{eff}}^{\,2}\,\delta\omega_{\rm GR}^2}, \hspace{3mm} \mbox{with} \hspace{3mm}
\delta\omega_{\rm GR} = \omega_0 \frac{g\,h}{c^2}
\label{Fisher02}
\end{equation}
This equation illustrates how nuclear heterodyne experiments can be scaled to reach 
desired precision values for Mössbauer isotopes of interest. 
The dimensionless prefactor $\beta$ in Eq.~(9) arises from the exact evaluation of the Fisher integral over the finite observation window. It accounts for the cycle-averaged Fisher weight of the heterodyne oscillation and for the fact that the rate-weighted mean of $t^2$ is smaller than $t_{\mathrm{eff}}^2$, because most delayed photons arrive at early times where $\delta\omega\,t$ is still small. For the $^{57}$Fe benchmark parameters with an observation window of $[10,\,200]$\,ns, one obtains $\beta = 0.234$; see \cite{Supplemental}, Sec.~V. Equation~(9) is further supported by Monte Carlo simulations in which synthetic Poisson-sampled heterodyne waveforms are processed through the full analysis chain, confirming that the estimator is unbiased and statistically consistent with the Fisher-information analysis for the benchmark cases considered; see \cite{Supplemental}, Sec.~IX.

The Fisher-information formalism can be extended to several fit parameters at once, so that timing offsets, 
detector jitter, heterodyne detuning, and normalization factors are treated together 
with the gravitational frequency shift. The inverse Fisher matrix then provides not only the
minimum uncertainty of the GRS estimate, but also quantitative requirements on timing stability, detuning control, and normalization accuracy, see
\cite{Supplemental}, Sec.\,VII.A.

\textit{Experimental feasibility} --
All key components required for nuclear heterodyne spectroscopy are either available at present synchrotron beamlines for nuclear resonant scattering or can be prepared with established techniques in our labs. The resonant absorbers consist of stainless steel Fe$_{55}$Cr$_{25}$Ni$_{20}$ foils enriched to $>95\%$ in $^{57}$Fe, which act as single-line scatterers \cite{Velten2024}. The effective linewidth $\Gamma_{\rm eff} \approx \Gamma_0(1 + T_M/4)$, where $T_M$ is the Mössbauer effective thickness, is determined by the foil thickness and controls both the delayed count rate and the phase-accumulation time. The acquisition time is minimized by maximizing $R(T_M) \cdot t_{\rm eff}^2$, where $R \propto T_M^2$ (coherent enhancement) favors thick foils while $t_{\rm eff} = 2/\Gamma_{\rm eff}$ favors thin ones. Photoelectric absorption is negligible at the relevant thicknesses ($\alpha_{\rm pe} \approx 0.004$ per unit $T_M$), so the practical upper limit is set instead by the onset of dynamical beats in the nuclear forward scattering signal, which distort the waveform model for $T_M \gtrsim 5$. The optimal regime is $T_M \approx 2$--$5$ ($\Gamma_{\rm eff} \approx 1.5$--$2.25\,\Gamma_0$), corresponding to a physical thickness of $\sim$0.2--0.5\,$\mu$m for enriched stainless steel --- accessible via sputter deposition \cite{Sahoo2011}. The benchmark value $\Gamma_{\rm eff} = 2\Gamma_0$ ($T_M \approx 4$, $d \approx 0.4\,\mu$m) adopted in this work sits at the center of this range. 

Controlled heterodyne detuning is achieved using commercial Mössbauer velocity transducers driven with stabilized waveforms and monitored by laser Doppler interferometry. In the case of $^{57}$Fe, the beamsplitter will be realized as a few-10\,$\mu$m thin silicon crystal platelet \cite{Powers2025} using the well-established Si(840) Bragg reflection \cite{Siddons1993,Toellner1995,Rohlsberger1997,Heeg2013Vacuum,Heeg2015Tunable,Haber2016Collecti,Marx-Glowna2021} with a Bragg angle of $\theta_{\rm B} = 45.1^\circ$ at 14.4\,keV, and the thickness chosen to result in balanced intensities in both interferometer arms. Detection is performed with fast avalanche photodiodes offering sub-nanosecond timing resolution, low background ($<0.1~\mathrm{s^{-1}}$), and large dynamic range of up to $10^7$. The experimental hutch EH3 at beamline P01 of PETRA\,III (DESY, Hamburg) provides a vertical baseline of 4\,m, possibly extendable to 8\,m, sufficient for immediate implementation of first benchmark nuclear gravitational redshift measurements.

Resonant photon fluxes of order $5\times10^4~\mathrm{s^{-1}}$ have already been demonstrated at beamline P01 of PETRA~III under closely related conditions \cite{Velten2024}; the benchmark value $R\simeq2\times10^4~\mathrm{s^{-1}}$ per arm adopted here is a conservative estimate for the dual-arm geometry. With $C\simeq0.9$, $\Gamma=2\Gamma_0$ ($t_{\rm eff}\approx140\,\mathrm{ns}$), $\beta=0.234$, and a baseline of $h=4\,\mathrm{m}$, Eq.\,(\ref{Fisher02}) yields an acquisition time of order 2 hours for a $5\sigma$ observation of the gravitational redshift. Percent-level precision ($S=100$) is reached for an 8\,m baseline on timescales of about 8 days, compatible with realistic beamtime allocations; see Fig.\,1b.

\textit{Symmetry-based suppression of systematic effects} --
A useful way to assess the robustness of nuclear heterodyne interferometry is to classify both the gravitational signal and the dominant perturbations by their parity under three experimentally accessible operations: arm subtraction $L \leftrightarrow U$, heterodyne-detuning reversal $\Delta\omega_{\mathrm{het}} \to -\Delta\omega_{\mathrm{het}}$, and detector exchange. Common-mode perturbations such as synchrotron timing jitter, global energy drifts, and background offsets are even under arm subtraction and cancel to first order. Arm-dependent detuning or beamsplitter asymmetries are odd under detuning reversal and are removed by averaging over opposite detunings, while detector-related asymmetries are isolated by detector exchange. The gravitational redshift carries the unique parity signature $(\mathrm{odd},\,\mathrm{even},\,\mathrm{even})$ and is therefore the only relevant signal that survives all three operations; see also~\cite{Supplemental}, Sec.\,VII.\,F, Table~I. Taken together, these symmetry operations suppress the dominant instrumental effects to first order, so that the method can operate close to the counting-statistics-limited regime. The remaining systematics --- mainly residual arm-dependent spectral asymmetries, imperfect normalization, absorber temperature matching, and the absolute calibration of the baseline~$h$ --- enter only at reduced level but must still be controlled for sub-percent-level measurements.

\textit{Scalability with baseline, isotope, and source} --
A major advantage of nuclear heterodyne interferometry is its scalability with respect to baseline and isotope. Increasing the absorber separation from meter scale to the Pound--Rebka height difference of $h=22.6\,\mathrm{m}$, while keeping the same $^{57}$Fe benchmark parameters, enables a $5\sigma$ observation of the GRS in only $T_{5\sigma}\approx 2.5\,\mathrm{min}$. A 1\,$\%$ test of the deviation parameter $\alpha$ would require about $16\,\mathrm{h}$, while $2 \times 10^{-3}$ precision would be reached after about $17\,\mathrm{days}$ (assuming an observation window up to 400\,ns). This opens the prospect of improving the historical nuclear-sector benchmark of $\sigma_{\alpha}\sim10^{-2}$ by about a factor of  5. Conversely, meter-scale baselines already permit measurements of the GRS as a function of $h$ within the same apparatus, testing the expected scaling $\Delta E/E=gh/c^2$ and improving the calibration of residual systematics. The method is furthermore extendable to other Mössbauer isotopes, including $^{181}$Ta and $^{119}$Sn, because the relevant resonant transitions are driven directly from the ground state and do not require radioactive sources. The concept can be reproduced at suitable synchrotron facilities worldwide. Future implementations at diffraction-limited sources and X-ray lasers are expected to provide further gains in spectral flux, timing structure, beam quality, and coherence, with corresponding improvements in sensitivity and acquisition time.

\textit{Conclusion and Outlook} --
Conceptually, nuclear heterodyne interferometry can be viewed as the nuclear analogue of Ramsey interferometry realized in the time domain of nuclear resonant scattering, exploiting nuclear coherence in a manner closely related to atomic clocks.  After pulsed resonant excitation, the nuclei in the two absorbers evolve as coherent oscillators at different gravitational potentials, analogous to clocks operating at slightly different frequencies. In the present scheme, the heterodyne reference converts the nuclear response into beat signals, while the gravitational redshift manifests as a slowly accumulating phase difference. The analogy to Ramsey spectroscopy is well founded at the level of phase accumulation, but the readout here is based on subtraction of intensities rather than quantum amplitude recombination. This is a practical advantage, removing the need for phase-stable path recombination and rendering the method robust against path-length fluctuations. 
The experimentally confirmed count rates and Monte Carlo validation of the full analysis chain — spanning two orders of magnitude in count rate, a factor of 300 in coherence time, and baselines from 4 to 22.6\,m — establish the universality of the method across the nuclear chart. Extending it to isotopes like $^{181}$Ta, $^{119}$Sn, and to nuclear clock isomers such as $^{45}$Sc could therefore open nuclear heterodyne interferometry as a new experimental platform for precision tests of gravitational coupling to nuclear structure and the strong-interaction sector.


%

\vspace{5mm}

\begin{acknowledgments}
The author acknowledges helpful and stimulating discussions with Guido Meier, Sven Velten, J\"org Evers, Joachim von Zanthier, Thomas St\"ohlker and Wolfgang Schleich
\end{acknowledgments}

The Monte Carlo simulation code and data that support the findings of this article are available from the author upon reasonable request.

\end{document}